\documentstyle[aps,epsfig,pre,epsf]{revtex}
\def\inseps#1#2{\def\epsfsize##1##2{#2##1} \centerline{\epsfbox{#1}}}
\begin{document}
\draft
\twocolumn[\hsize\textwidth\columnwidth\hsize\csname 
  @twocolumnfalse\endcsname]
\title{Rheological Behavior  of Microemulsions}
\author{G. Gonnella and M. Ruggieri}
\address{Istituto Nazionale per la  Fisica della Materia, Unit\`a di Bari
{\rm and} Dipartimento di Fisica, Universit\`a di Bari, {\rm and}
Istituto Nazionale di Fisica Nucleare, Sezione di Bari, via Amendola
173, 70126 Bari, Italy.}
\date{\today}
\maketitle
\begin{abstract}
We study the stationary  and transient behaviors of the microemulsion 
phase subjected to a shear flow.
The system is described by a diffusion-convective equation
which generalizes the usual Cahn-Hilliard
 equation. 
Non-linear terms are treated in a self-consistent
approximation.
Shear, first and second normal stresses are calculated
as momenta of the structure factor.
Shear thinning is observed in  stationary conditions.
After a newtonian regime at small values of the shear
rate, the excess viscosity decreases when the shear rate 
becomes  of the order of the inverse of the relaxation time
of the system without flow. In transient regimes,
when the flow is applied starting from a quiescent 
state, we find that the shear stress reaches a maximum before
decreasing to a constant value.

\end{abstract}

\section{Introduction}

The rheological behavior of complex fluids such as polymer solutions,
polymer melts, emulsions is of considerable interest both in technology
and in basic research~\cite{larson}.
While the behavior of the stress response to  applied flows 
is of fundamental importance in many applications, it also reflects
the existence of mesoscopic structures in the fluid and 
is intimately related to its constitution.
For example, when a shear flow is applied to a polymer solution, the stress
first reaches a maximum and then relaxes to a constant value~\cite{LJ73}.
This phenomenon, at small strain, 
 is related to the  entanglement
of the polymer network which is   distorted by the flow
with a resulting 
 increase of the  stress. At larger values of the strain, 
however,  the  disentanglement of 
the system is favoured and the stress is observed to decrease.
In general, non monotonic relaxational properties of the stress are
typical of  complex fluids which are also characterized,
in stationary conditions,  
by  non-newtonian behavior. The effective viscosity depends
on the applied shear flow and different 
behaviors can be   observed~\cite{larson}.

In this paper we consider the rheological behavior 
of the microemulsion phase  both in stationary 
and transient conditions. In ternary self-assembling systems
the surfactant 
forms  interfaces between  oil- and water-like 	domains.
These interfaces, in  the microemulsion phase, constitute an intertwined 
bicontinuous structure disordered on large scales but with mesoscopic 
order on distances of the order of 500  Angstroms \cite{GR}.
The observed  structure factor is given by 
\begin{equation}
I(q) \sim \frac{1}{a+g q^{2}+c q^{4}} 
\label{eqn1}
\end{equation}
which,  for $g <  0$,  has a maximum at 
$q=\sqrt{\frac{|g|}{2c}}$\cite{TS,W,col}.
In real space, this corresponds to the two-point correlation function 
\begin{equation}
G(r)  =  \frac{d}{2\pi r}e^{-r/\xi}\sin(\frac{2\pi r}{d})
\label{eqn2}
\end{equation}
where $\xi$ plays the role of the usual correlation length
in disordered phases, $d$ is related to the size of coherent regions
of oil or water domains, and
 typical values of the ratio $d/\xi$ are in  the interval $2-4$.
 
The behavior of the stress in the microemulsion phase
was first considered in \cite{Daw1} and then in \cite{Sup} where also the 
two-time correlation functions were studied. 
Here we complete the analysis of the steady state of \cite{Sup} 
and consider also  the transient behavior. Our approach is based 
on the use of a continuum free-energy
functional and is similar to that of
Onuki and Kawasaki \cite{OK}, applied  also  
 to evaluate the effects of a shear flow
on copolymer melts\cite{Fre,FL},
 on the
 disorder-lamellar transition \cite{cates,HM},
 and on the phase separation of binary mixtures \cite{noi}.

We consider  a Cahn-Hilliard equation generalized
by the presence of a convective term. Hydrodynamical effects are
 neglected; moreover, 
the surfactant  is assumed to relaxe faster than 
the other components of the mixture so that its degrees of freedom
are not explicitely  considered.
Non-linear terms, 
which become relevant close to transition lines, will be treated  
self-consistently.
A renormalization procedure is introduced and 
the system is studied in terms of the  physical 
variables $\xi,d$ of  the case without flow. 

Our main result for the stationary regime  is the behavior 
of the constitutive curve.
Shear thinning, which is the decreasing of the effective viscosity
when the  shear rate is increased, 
is observed in   two different  ranges of the shear rate.
It  first occurs at a value of the
shear rate of the order of the inverse of the relaxation time
of microemulsions without shear. 
The  morphological changes 
occurring   when the shear rate is increased can be deduced 
by the patterns exhibited by the structure factor.
In the transient behavior  after the application of the flow, 
at sufficiently high shear rates,
we observe a maximum in  the shear stress followed by 
a relaxation to a constant value, analogously to what observed 
in other systems. We have also studied the  behavior of the
stress  tensor when, starting from a stationary state with  shear, 
the flow is switched off. 

The paper is divided as follows.
In Section 2 we specify the model and solve formally the 
dynamical equation for the structure factor.
Results for  the stationary  regime and for the transients are 
respectively 
described in  Sections 3 and 4.
Section 5 contains some conclusions.

\section{The model}

Our  study of  rheological properties  of microemulsions
 is based on a  functional Landau-Ginzburg approach with one 
scalar order parameter  $\phi(\vec{x})$  representing 
the concentration difference between oil and water. We  consider the
hamiltonian
\begin{eqnarray}
{\cal H}[\phi]&=&\int d^{3}x\,\left\{
    \frac{1}{2}\left[a_{2}\phi^{2}+(g_{0}+g_{2}\phi^{2})\left(
      \vec{\nabla}\phi\right)^{2}\right.\right.\nonumber\\
        &+&\left.\left. c\left(\nabla^{2}\phi\right)^{2}\right]
   +\frac{\lambda}{4}\phi^4 \right\}
\label{eq1}
\end{eqnarray}
which has been largely demonstrated to well describe equilibrium properties 
of ternary mixtures~\cite{GR}. 
Here we briefly discuss the  properties of this hamiltonian
relevant for the microemulsion phase.
The expression (\ref{eq1}) differs  in  the gradient terms
from the usual Landau-Ginzburg 
hamiltonian used to study
 binary mixtures. A negative value
of the function  $g(\phi) = g_{0} + g_2 \varphi^2$ ($g_2 > 0 $)
favors the appearing of interfaces. In particular the value    
of $g_{0}$ can be related to the amount of surfactant present in the system.
The term proportional to $c>0$ assures stability  at large momenta 
and weights the curvature of interfaces. 

The presence of the quartic terms $\phi^{4}$ and 
 $\phi^{2}\left(\vec{\nabla}\phi\right)^{2}$,
which also in a disordered phase could have a role in the proximity of
a transition line,   
 makes impossible an exact determination of the
two point correlation functions. However, following
~\cite{Daw1}, it is possible 
to use a renormalization procedure based on  a 
self-consistent
approximation to find an expression for the equilibrium scattering function
and  for the two-point correlation function in
real space. For the first, defined $\phi(\vec{k})$ the Fourier transform of
$\phi(\vec{x})$, it is found that
\begin{equation}
S(k)\equiv<\phi(\vec{k})\phi(-\vec{k})>=
 \frac{T}{a_{r}+g_{r}k^{2}+ck^{4}}
\label{eq2}
\end{equation}
where $k\equiv|\vec{k}|$, the renormalized parameters are given by
\begin{eqnarray}
a_{r}&=&a_{2}+ \lambda {\cal S}_{0}+g_{2}{\cal S}_{2} 
\label{eq3} \\
g_{r}&=&g_{0}+g_{2}{\cal S}_{0}
\label{eq4}
\end{eqnarray}
 and the loop integrals are defined as
\begin{equation}
{\cal S}_{p}=\int_{|\vec{k}|<\Lambda}\frac{d^{3}\vec{k}}{(2\pi)^{3}}\,
   k^{p}S(k)
\label{eq5}
\end{equation}
($p=0,2$) with  $\Lambda$ being 
 a high momentum phenomenological cut-off. The 
region $g_{r}<0$ and $4a_{r}c-g_{r}^{2}>0$, as discussed in the
introduction,
can be  identified with the microemulsion phase with 
the function $S(k)$ having a peak at $k\equiv k_{M}=\sqrt{\frac{|g_{r}|}{2c}}$.
Moreover, 
the characteristic lenghts $\xi$ and $d$ appearing in the 
real space two-point correlation
function (\ref{eqn2}) are given by
\begin{eqnarray}
\xi&=&\left[\frac{1}{2}\left(\frac{a_{r}}{c}\right)^{1/2}
+\frac{1}{4}\left(\frac{g_{r}}{c}\right)\right]^{\frac{1}{2}}
\label{eq7} \\
d&=& 2\pi\left[\frac{1}{2}\left(\frac{a_{r}}{c}\right)^{1/2}
-\frac{1}{4}\left(\frac{g_{r}}{c}\right)\right]^{\frac{1}{2}} \qquad .
\label{eq8}       
\end{eqnarray}
Therefore  equilibrium properties  can be expressed in terms 
  of renormalized parameters and, by equations (\ref{eq7}) and (\ref{eq8}), 
through  the physical lenghts $\xi$ and $d$ 
(once the parameters $\lambda$, $c$, 
$g_{2}$ and the cut-off $\Lambda$ are given).
\vskip0.3cm
The dynamics of the order parameter 
in presence of convective motion \cite{On97} is described  by the equation:
\begin{equation}
\frac{\partial\,\phi}{\partial\,t}
   +\vec{v}\cdot\vec{\nabla}\phi=
   \Gamma\nabla^{2}\left(\frac{\delta {\cal H}}{\delta\,\phi}\right)
   +\eta_{\phi} 
\label{eq9}
\end{equation}
where ${\cal H}$ is the hamiltonian of  (\ref{eq1}). 
The velocity field  $\vec{v}$ is  a planar Couette shear flow:
\begin{equation} 
\vec{v}=\gamma y\vec{e}_{x}
\label{eq10}
\end{equation}
where $\gamma$ is the shear rate and $\vec{e}_{x}$ the unit vector in 
the flow direction; $\eta_{\phi}$ is a white gaussian noise representing 
thermal fluctuations with momenta  given by
\begin{eqnarray}
<\eta_{\phi}(\vec{x},t)>&=&0
\label{eq11a} \\
<\eta_{\phi}(\vec{x},t)\eta_{\phi}(\vec{x}',t')>&=&
     -2T\Gamma\nabla^{2}\delta^{3}(\vec{x}-\vec{x}')\delta(t-t')
\label{eq11b}
\end{eqnarray}
($<\cdots>$ means the ensemble average) as required by the  
Fluctuation-Dissipation theorem which holds in absence of flow.
 The functional derivative $\delta{\cal H}/\delta\,\phi$ represents the
difference in chemical potentials between oil and water; $\Gamma$ is a mobility
coefficient and $T$ is the temperature of the heat bath. By assuming equation
(\ref{eq9}) as the evolution equation for $\phi$, we are neglecting
hydrodynamics fluctuations as well as the motion of the surfactant.

We will study the  evolution equation for the dynamical 
structure factor
\begin{equation}  
S(\vec{k},t) \equiv <\phi(\vec{k},t)\phi(-\vec{k},t)>
\label{eq_fattdin}
\end{equation}
in the same  self-consistent approximation used in equilibrium
to write Eqs.~(\ref{eq3},{\ref{eq4}). The  convection-diffusion  equation
 can be  formally linearized as \cite{nota}  
\begin{eqnarray}
\frac{\partial\,\phi}{\partial\,t}
   +\vec{v}\cdot\vec{\nabla}\phi&=&\Gamma\,\nabla^{2}
     \left\{\left(a_{2}+ \lambda{\cal S}_{0}(t)
       +g_{2}{\cal S}_{2}(t)\right)\phi
         \right.     
  \nonumber\\
&-&\left.(g_{0}+g_{2}{\cal S}_{0}(t))\nabla^2\phi + c\,\Delta^{2}\phi\right\}
+\eta_{\phi}
\label{eq_convdifflin}
\end{eqnarray}
where the quantities ${\cal S}_{p}(t) $  are  
given by   expressions analogue to those of  Eq.~(\ref{eq5}) 
but now with $ S(\vec{k},t)$ of Eq.~(\ref{eq_fattdin})
self-consistently calculated with Eq.~(\ref{eq_convdifflin}).
A standard procedure gives from Eq.~(\ref{eq_convdifflin}) the dynamical
 equation for $S(\vec{k},t)$: 
\begin{equation}
\left\{\frac{\partial}{\partial\,t}-\gamma
k_{x}\frac{\partial}{\partial\,k_{y}}
   +2\Gamma k^{2}K_{R}(k)\right\}S(\vec{k},t)=2T\Gamma k^{2}       
\label{eq12}
\end{equation}
where $K_{R}(k)\equiv a_{r}+g_{r}k^{2}+ck^{4}$ is the 
renormalized vertex
function and the parameters $a_{r}$ and $g_{r}$ can be obtained
as in Eqs.~
(\ref{eq3},\ref{eq4}) using ${\cal S}_{p}(t) $.

A formal  solution of Eq.~(\ref{eq12}) may be obtained by  the 
method of characteristics:
\begin{equation}
S(\vec{k},t)=\Delta_{0}(\vec{{\cal K}}(t))\,{\cal I}_{1}(t)
 + 2\,T\,\Gamma\,{\cal I}_{2}(t)    
\label{eq13}
\end{equation}
where we have defined the functions
\begin{eqnarray}
{\cal I}_{1}(t)&=&
 e^{-2\Gamma\int_{0}^{t}ds\,{\cal K}^{2}(s)
[a_{r}+g_{r}{\cal K}^{2}(s)+c{\cal K}^{4}(s)]} 
\label{eq_def1} \\
{\cal I}_{2}(t)&=& \int_{0}^{t}du\,{\cal K}^{2}(u)\,{\cal I}_{1}(u)
\label{eq_def2}
\end{eqnarray}
and $\vec{{\cal K}}(u)\equiv\vec{k}+\gamma
k_{x} u \vec{e}_{y}$; $\Delta_{0}(\vec k)$ is the
 structure factor at the initial time $t=0$.
Since the quantities $a_{r}$ and $ g_{r}$ 
contains the momenta of $S(\vec{k},t)$,  Equation (\ref{eq13})
is actually a nonlinear integral equation
for  $S(\vec{k},t)$. This equation 
can be solved numerically
at each  time by iterative methods.

Our results will first concern  steady state  properties. 
The stationary
solution can be  readily obtained from the  $t\rightarrow +\infty$ limit
of Eq.~(\ref{eq13}),
 observing that in this limit the first term
of the solution tends to zero (except for  the $\vec{k}=0$ mode). 
Therefore we write
the stationary structure factor as:
\begin{equation}
S(\vec{k};\gamma)_{\infty}=2\,T\,\Gamma\,{\cal I}_{2}(\infty)
\label{eq14}
\end{equation}
where
\begin{eqnarray}
{\cal I}_{2}(\infty)&=&\int_{0}^{\infty}dz\,{\cal K}^{2}(z)\,
 e^{-2\Gamma\int_{0}^{z}ds\,{\cal K}^{2}(s)
[a_{r}+g_{r}{\cal K}^{2}(s)+c{\cal K}^{4}(s)]}\nonumber
\label{eq14_bis}
\end{eqnarray}

We will also study transient  behaviors with the fluid  evolving
 from a quiescent state  towards the stationary state with  shear, 
 or with the system relaxing,  after interruption of the  flow,
from the   sheared stationary state  into the quiescent state.
For the latter case  we use the solution of 
equation (\ref{eq12}) with $\gamma =0$:
\begin{eqnarray}
S(\vec{k};\gamma;t)_{relax}&=&S(\vec{k};\gamma)_{\infty}
   e^{-2\Gamma k^{2}K_{R}(k)} \nonumber\\
&+&\frac{T}{K_{R}(k)}\left(1-e^{-\frac{t}{2\Gamma k^{2}K_{R}(k)}}\right)   
\label{eq15}
\end{eqnarray}
 
Finally, once the structure factor is known, we may evaluate the stresses
which  can be obtained as momenta 
of the structure factor \cite{Daw1}.
The shear, first and second normal stresses are 
respectively given by
\begin{eqnarray}
\sigma_{xy}(t)=-\int_{|\vec{k}|<\Lambda}\frac{d^{3}\vec{k}}{(2\pi)^{3}}\,
   k_{x}k_{y}(g_{r}+2ck^{2})S(\vec{k},t) 
   \label{eq16a} \\
N_{1}(t)=-\int_{|\vec{k}|<\Lambda}\frac{d^{3}\vec{k}}{(2\pi)^{3}}\,
   (k_{x}^{2}-k_{y}^{2})(g_{r}+2ck^{2})S(\vec{k},t) 
   \label{eq16b} \\   
N_{2}(t)=-\int_{|\vec{k}|<\Lambda}\frac{d^{3}\vec{k}}{(2\pi)^{3}}\,
   (k_{y}^{2}-k_{z}^{2})(g_{r}+2ck^{2})S(\vec{k},t) 
   \label{eq16c}   
\end{eqnarray}
In addition, the excess viscosity is defined as:
\begin{equation}
\Delta\eta(t)=\frac{\sigma_{xy}}{\gamma} 
\label{eq17}
\end{equation}
which represents the contribution of interfaces to the 
full viscosity of the fluid (that is, evaluating the viscosity
of the fluid by means of equation (\ref{eq17}) we are neglecting
the hydrodynamical contribution to the viscosity itself).

\section{Stationary regime}

In this section we present   results for  the steady 
states reached under the action of the shear flow with the 
structure factor given by  Eq.~(\ref{eq14}). 
We have studied numerically this expression 
for several values of $\xi, d$ and $\gamma$. 
The other parameters have been fixed as $g_2= 1, c = 1,
\lambda = 0.5, \Lambda = 3$.

The effects of the flow on the structure factor can be seen
in Fig.~1 where the
projections  on the planes
$k_y=0$ and $k_z=0$ are shown for  different $\gamma$ and $\xi=2, d=6$. 
Similar results have been obtained for other choices
of $\xi$ and $d$ \cite{Sup}.
(At $k_x=0$ 
the shape of the structure factor is the same of the  case without
 flow, see Eq.~(\ref{eq12})).
At $\gamma = 0.5 $ the structure factor remains almost isotropic 
and its pattern
for  each cartesian
plane is close to that  of a circular volcano.
The patterns  are 
progressively distorted when the shear rate is increased.
On the plane  $k_z = 0 $, at  $\gamma = 2 $, 
 the edge of the volcano has assumed an  elliptical shape 
and   four peaks are visible.
These peaks initially appear
on the coordinate axes;
then, when $\gamma $ is 
 increased,   the ones located at 
$k_x\simeq 0$ become comparatively more important while the two others
rotate clockwise and decrease their amplitude linearly with $\gamma $
until they disappear.  
Indeed, in the limit $\gamma \to \infty$, 
since terms proportional to powers of
$\gamma k_x$ damp the exponential term on the r.h.s. of Eq.~(\ref{eq14}),
only the maxima of $C(\vec k)$ with $k_x=0$ and  $k_y  = \pm k_M$
survive.  On the other  plane
$ k_y = 0$ 
two peaks at $k_x = 0, k_z = \pm k_M $, 
 are also observed to become sharper and sharper as $\gamma$
is increased. 

The above results  can be related to the orientation of the interfaces 
in the mixture, as also  observed in \cite{Sup}.
A peak of $C(k)$  defines 
a characteristic length  proportional to the inverse
of its position and,
 since the system is not isotropic,
to each maximum one associates three lengths, one for each space direction.
Due to the symmetry $\vec k\to -\vec k$ only the peaks  
 not related by reflection around the origin can be considered.
At very  large shear rate the existence of  a single couple of  maxima
at $k_x=0$ signals that interfaces are preferentially aligned along the flow
with symmetry recovered 
in the transverse directions 
and the characteristic lengths being the same as without shear.
For intermediate values of $\gamma $ 
the additional peaks at  $(\tilde k_x, \tilde k_y, \tilde k_z)$ 
reveal the 
presence of interfaces oriented with an angle 
$\alpha = \arctan (-\tilde k_x/ \tilde k_y)$ with respect to the flow, besides
those aligned along the $x$ direction. These additional peaks are better
seen in a region of parameters 
closer to the microemulsion-lamellar
transition line corresponding to  a larger value of $\xi$~\cite{Sup}.
As $\gamma $ is
increased the tilt angle $\alpha$ and the  relative abundance of lamellae
orienteted  at this angle
diminish
as suggested by the behavior of the maxima with $k_x\ne 0$
previously discussed. 

The  behavior of the stress tensor  as a function of the shear rate
is  reported in Figs.~2-4. At small values of $\gamma$
the shear stress is a linear function of $\gamma$  
so that the viscosity is constant and
 the fluid is newtonian. Shear thinning occurs 
for $\gamma $ between 1 and 2, 
 when the slope of the curve of the stress
changes significantly. At this point the original volcano shape of
the structure factor has also appreciably changed. 
In terms of interfaces we expect that, when shear 
thinning is observed, the bicontinuos network of interfaces
which is only distorted in the newtonian regime, is affected by many 
 ruptures with a significant dicrease  of 
connectivity.  We can say that a strong shear regime is entered. Indeed, 
 we can compare 
the shear temporal scale $\gamma^{-1}$  with the  relaxation time $\tau_M$
of microemulsions in equilibrium.
The relaxation time of a mode with wavevector $k$ is given by 
\begin{equation} 
\tau(\vec{k})=\frac{1}{\Gamma {k}^{2}
   (a_{r}+g_{r} {k}^{2}+c {k}^{4})}
\label{eq_dec_modok}
\end{equation}
Following ~\cite{Daw1} we choose $k=k_M$ corresponding to the peak of 
microemulsions, so that
\begin{equation}
\tau_M =\frac{1}{8 \Gamma c}
   \frac{\xi^{2}d^{2}}{4\pi^{2}[(2\pi^{2}/d)^{2}-1/\xi^{2}]} \quad .
 \label{eq_tm}
\end{equation}
The corresponding Deborah number is given by 
\begin{equation}
De =\frac{\tau_M}{\tau_S} 
\label{deborah}
\end{equation}
where $\tau_S  = 1/\gamma$.
If we take  $\gamma = 2$  we get $De \sim 1.9 $ for
 the case of Fig.~2. This  indicates  that 
shear thinning becomes evident when the shear rate is of the order
of the inverse of typical structural times of the system without
shear.  
We checked for other values of $\xi,d$
that the Deborah number at shear thinning  is always of  order  1.

Shear thinning  is observed again at $\gamma \sim 10^3$
 when, as we have seen, 
peaks with $k_x =0$ largely prevail. 
We expect that at  these values  of the shear rate the original 
bicontinuous interface network has changed significantly
its topology becoming more similar to a stack of lamellae.
 At very large $\gamma$ 
the excess viscosity is found to decrease as 
$1/\gamma^{- s}$ with $s = 1.87 $ which
 is close to the analytical limit 
$s = 2  $ \cite{Daw1}. In this 
final stress regime  the lamellae are expected to become
more and more aligned with the flow with fluctuations
very inhibited.
We observe that the shear stress corresponding to 
a completely ordered lamellar phase is zero. 

The other stress components $N_1, N_2$
behave similarly. At small $\gamma$, $N_1, N_2   \sim \gamma^{2} $ 
while they decrease as $\gamma^{-1}$
 when $\gamma \rightarrow \infty$ (see Figs.~3-4).

\section{Transients}

We have studied the evolution 
of the system  under the action of the   shear flow
from the  initial  equilibrium 
configuration  of Eq.(\ref{eq2}) towards the steady state of the previous section,
as described by Eq.~(\ref{eq13}).
The behavior of the   stress components for different values of $\gamma$ 
is shown in Figs.~5-7. 
When  $\gamma$ is large enough that $ De \ge 1$,
 a non monotonic behavior of the stress is  observed
with $\sigma_{xy}, N_1, |N_2 |$ exhibiting a maximum
before relaxing to a constant value.
A similar  behavior has been measured
in polymer solutions\cite{LJ73}. 
In our case  we can think
that at initial times the surfactant interfaces are stretched by the flow
with a consequent increase of the stress. When the maximum of the stress
is reached, 
the interface structure  starts to be broken  and the stress relaxes
to a lower value.
The temporal evolution of the structure factor  for $\gamma = 100$
is shown 
in Fig.~8.
The largest observed  distortion   corresponds to 
the maximum of the stress. 
In the case $\gamma = 2 $, 
when the relaxation of the stress is monotonic,
a prolate pattern like that in the middle of Fig.8 at $k_z = 0 $
 is not observed.

We have also considered the opposite situation with the system,
initially in a stationary state with shear,
 evolving without  flow as described by Eq.~(\ref{eq15}).
In this case the behavior is exponentially monotonic 
after an initial faster decay, as it can be seen in Fig.~9.
The time constant $\tau$  of the exponential part 
of the relaxation decreases  with $\gamma$, as shown in Fig.~10.

\section{Conclusions}

We have used a generalized Cahn-Hilliard
equation with a convective term to study the rheological 
behavior of the microemulsion phase. 
The steady state constitutive curve shows shear thinning
  first occurring at a shear rate of the order of the inverse
of the equilibrium relaxation time.
We have also obtained for the first time analytical expressions
for the temporal behavior of the structure factor.
From this we derive 
a non monotonic evolution of the stress. This is 
 similar to what is observed  in other systems that  relaxe into
the steady state with a shear flow.
We hope that these predictions are useful for future experiments.
From the theoretical point of view this analysis can be completed
studying how the equilibrium phase diagram,
including the disorder and the Lifshitz lines \cite{GR,TS,W,col},
 is changed by the  presence
of the flow. Moreover, 
 hydrodynamic fluctuations should  be taken into account
for a full description of the system.

\acknowledgments
We thank Antonio Lamura and Federico Corberi for helpful discussions.
G.G acknowledges support by  PRA-HOP 1999 INFM.


\onecolumn
\newpage

\begin{center}
{\large FIGURES}
\end{center}

\begin{figure}
\begin{center}
\centerline{\epsfig{file=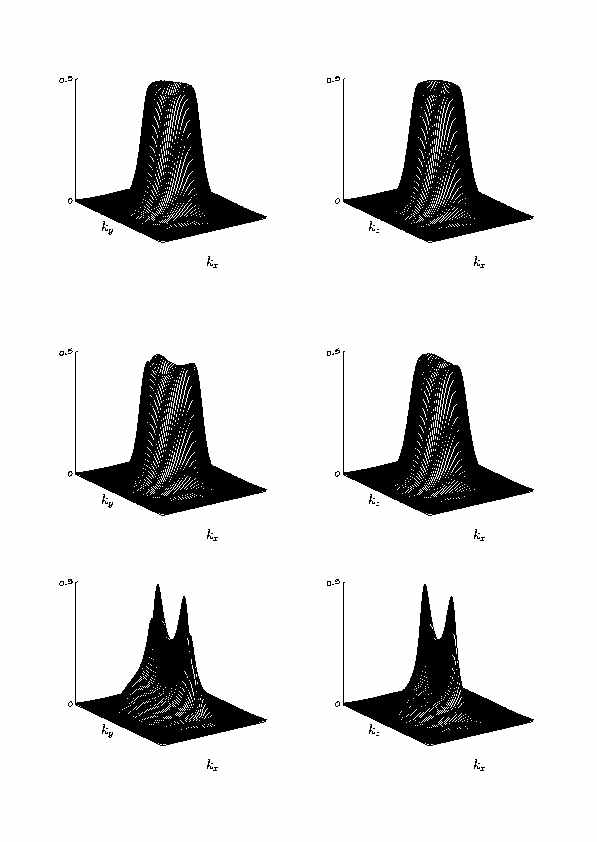,bbllx=55pt,bblly=30pt,bburx=535pt,bbury=850pt,
width=0.9\textwidth,height=0.85\textheight,clip=}}
\caption{Projections of the
structure factor in the stationary state 
on the planes $k_{y}=0$ (right column) and 
$k_{z}=0$ (left column) for
 $\xi=2$,  $d=6$. The shear rate is, from the top to the bottom,
 $\gamma=0.5$,
$\gamma=2$,  and $\gamma=100$; 
$k_{x}, k_{y}, k_z$ vary between
 $-3$ and $3$ in adimensional units.} 
\end{center}
\label{fig_1}
\end{figure}
\newpage

\begin{figure}
\begin{center}
\inseps{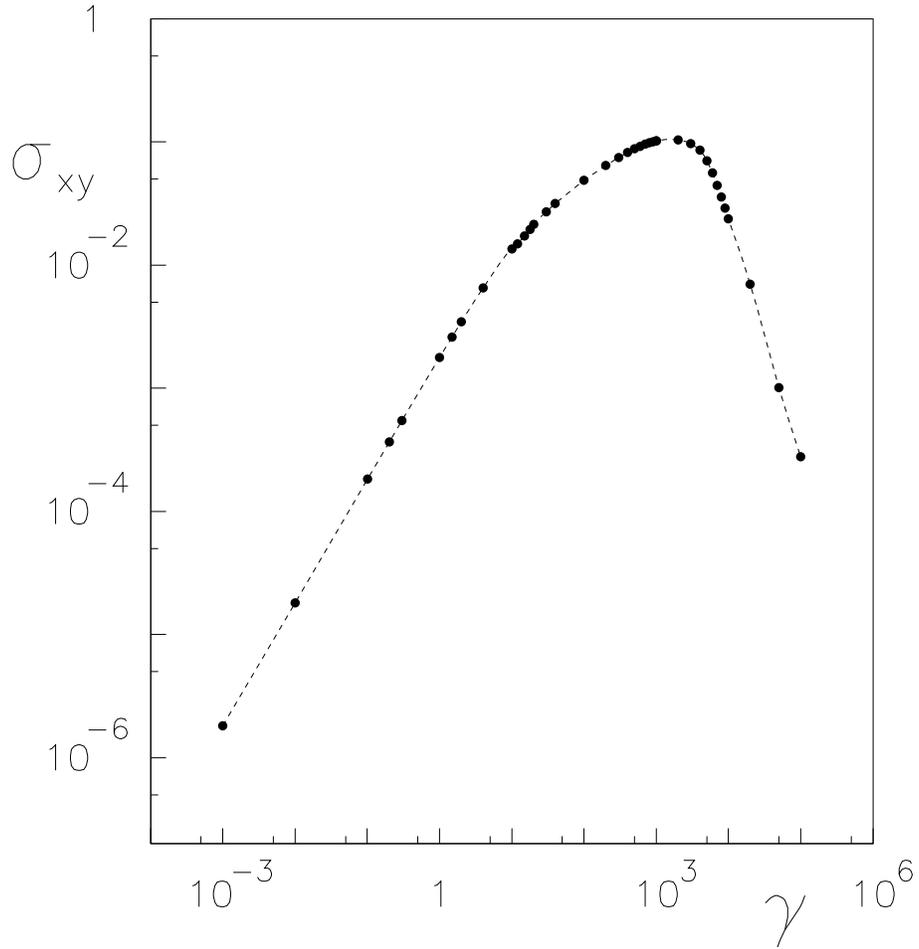}{0.7}
\end{center}
\caption{The stationary shear  stress $\sigma_{xy}$ versus 
the shear rate  $\gamma$ 
for  $\xi=2$ and $d=6$.}
\label{fig_3}
\end{figure}
\newpage

\begin{figure}
\begin{center}
\inseps{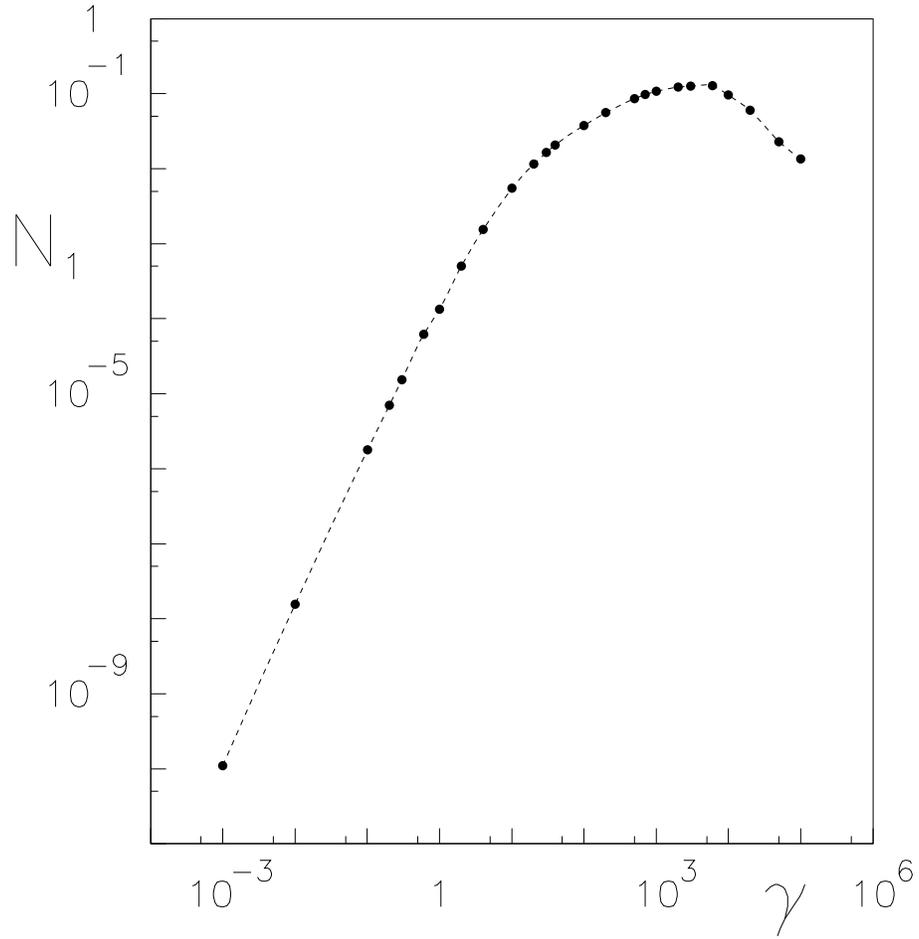}{0.7}
\end{center}
\caption{The first normal stress $N_{1}$  
 for $\xi=2$ and $d=6$.}
\label{fig_4}
\end{figure}
\newpage

\begin{figure}
\begin{center}
\inseps{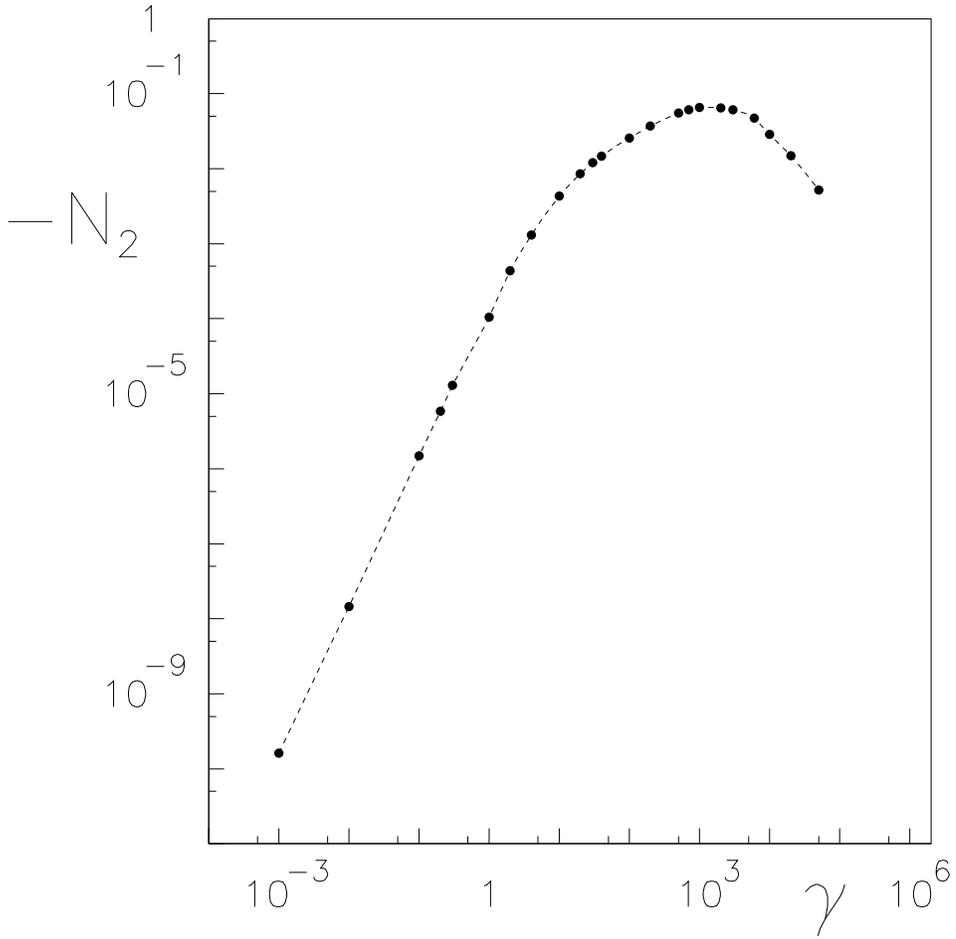}{0.7}
\end{center}
\caption{Absolute  value of  the  second normal stress $N_{2}$ 
 for $\xi=2$ and $d=6$.}
\label{fig_5}
\end{figure}
\newpage

\begin{figure}
\epsfig{file=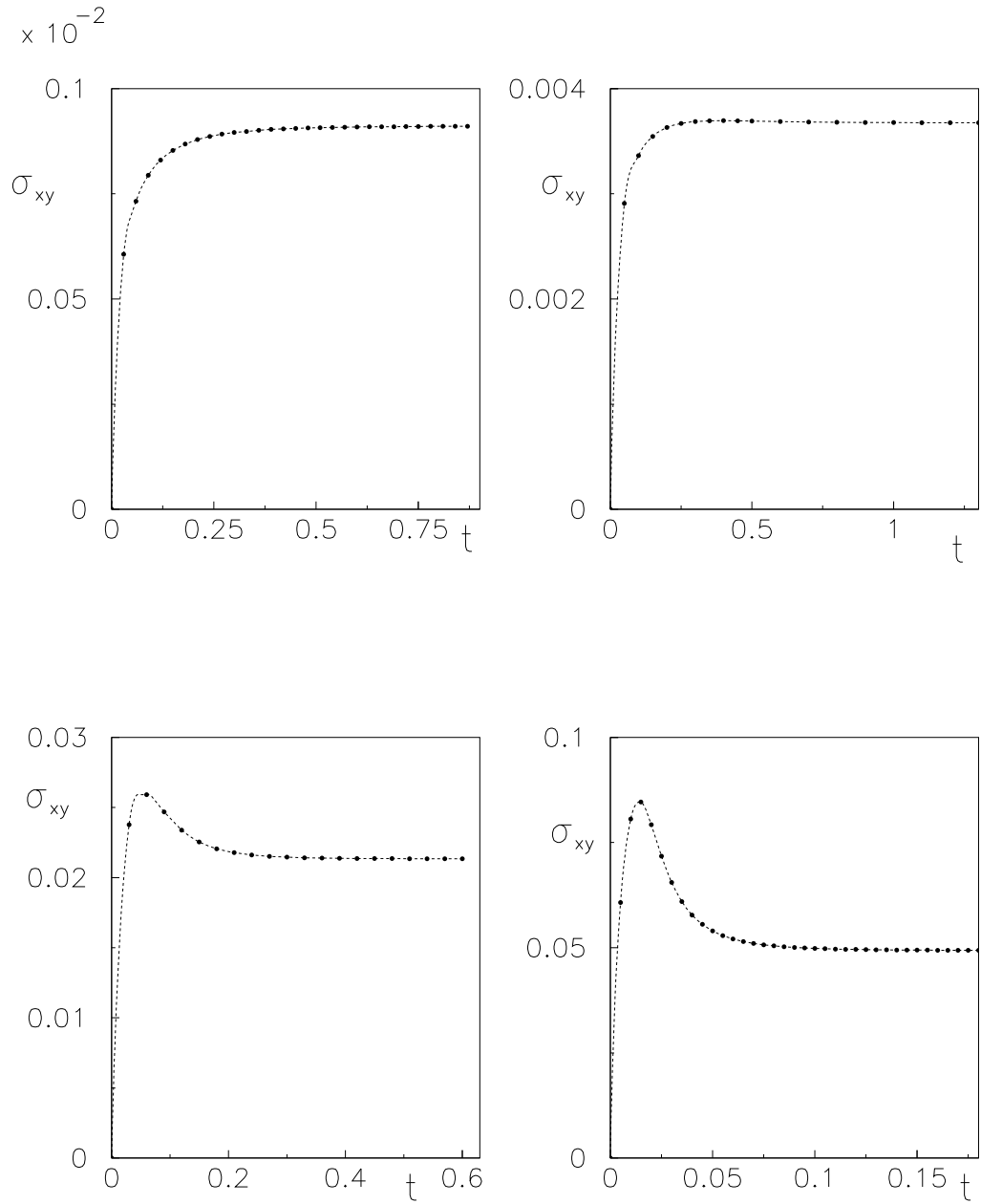,bbllx=70pt,bblly=221pt,bburx=509pt,bbury=700pt}
\label{fig_7}
\caption{Time evolution of the  stress $\sigma_{xy}$ 
for various values of the shear rate $\gamma$ with 
$\xi=2$ and $d=6$. Results are shown 
 for $\gamma=0.5$ (top-left), $\gamma=2$ (top-right),
$\gamma=20$ (bottom-left) and $\gamma=100$ (bottom-right).}
\end{figure}
\newpage

\begin{figure}
\epsfig{file=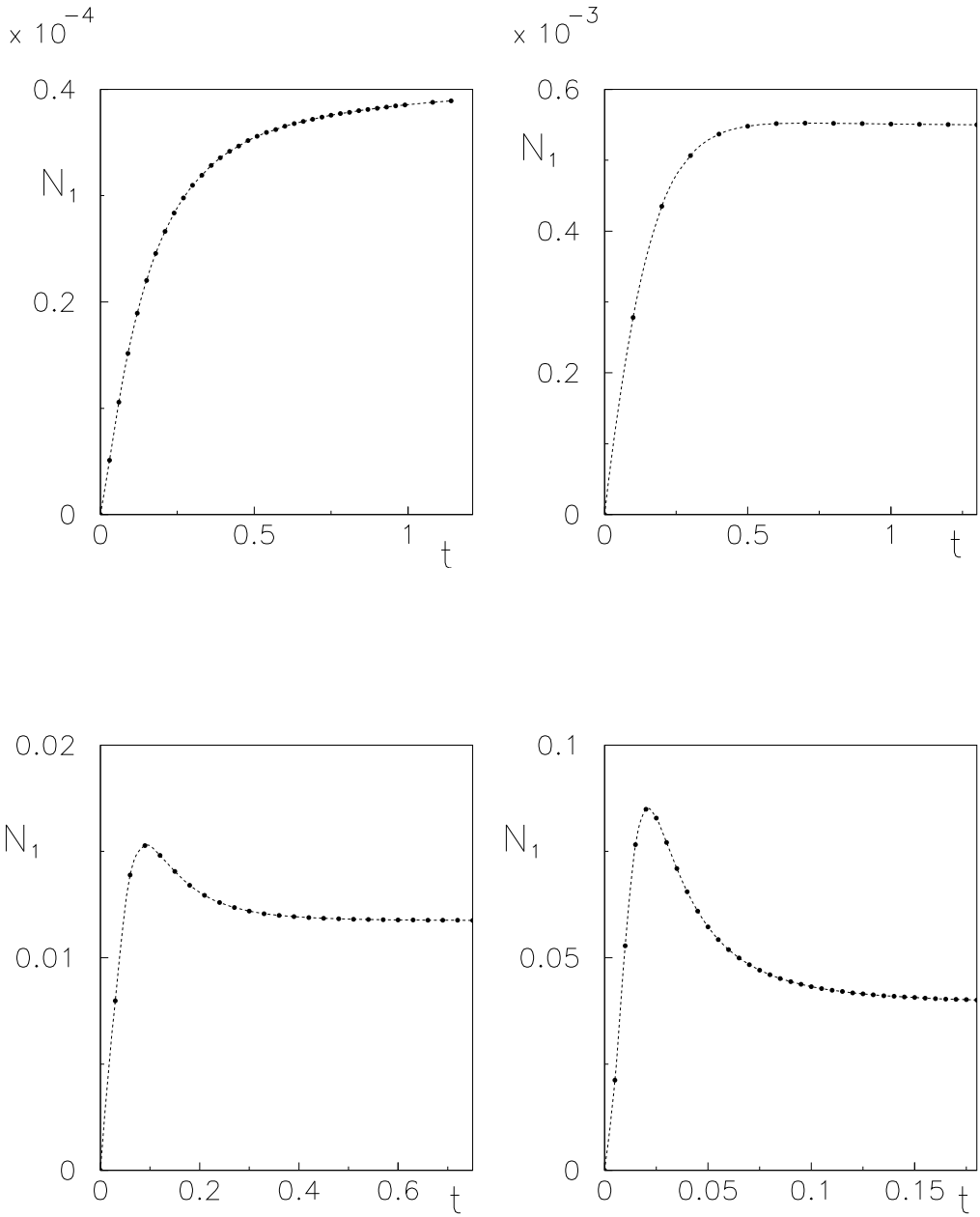,bbllx=90pt,bblly=221pt,bburx=509pt,bbury=700pt}
\label{fig_8}
\caption{Time evolution of the first normal stress $N_{1}$;
the parameters are the same of Fig.5.}
\end{figure}
\newpage

\begin{figure}
\epsfig{file=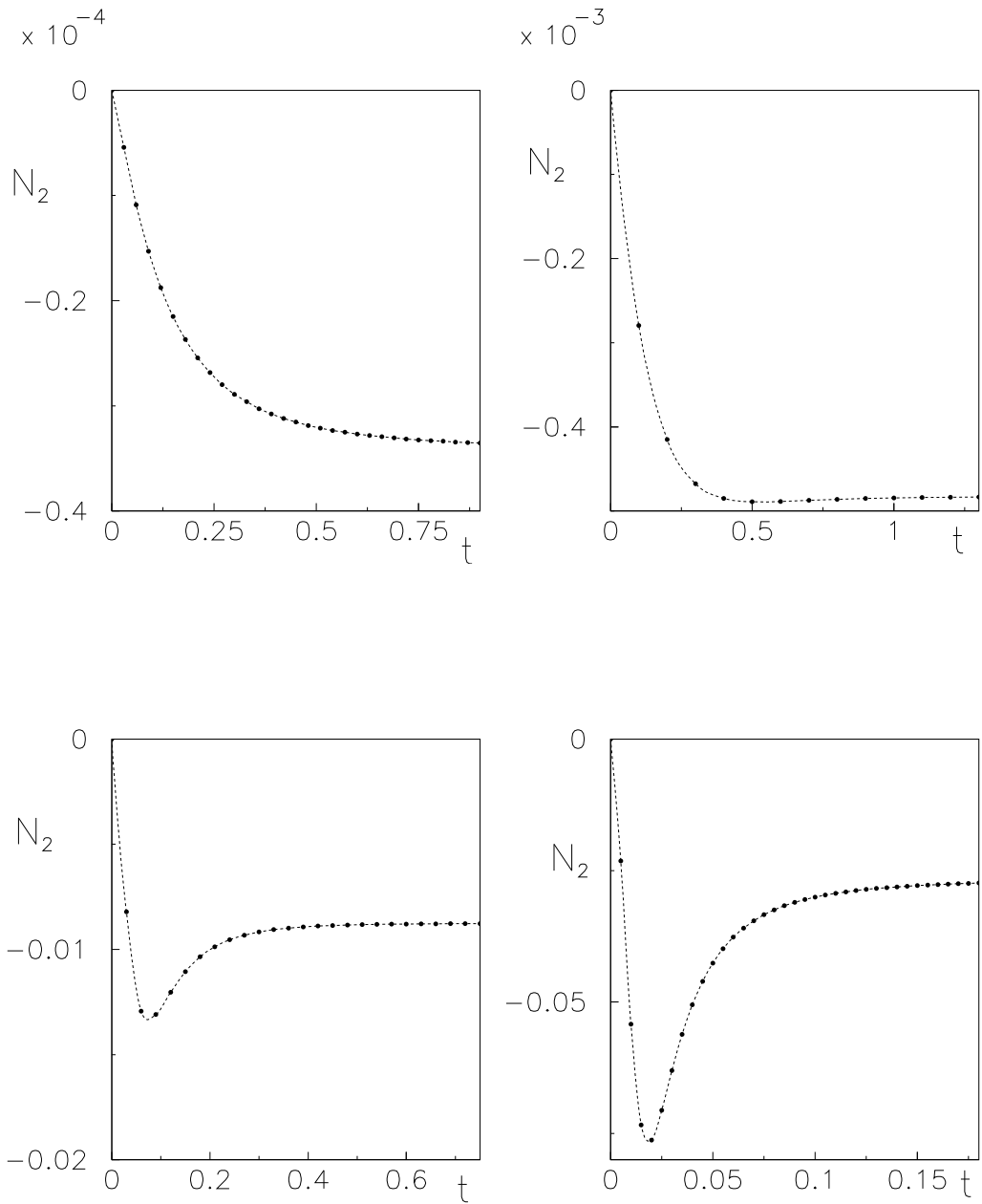,bbllx=90pt,bblly=221pt,bburx=509pt,bbury=700pt}
\label{fig_9}
\caption{Time evolution of the second normal stress $N_{2}$ with
 parameters  as in Fig.5.} 
\end{figure}
\newpage

\begin{figure}
\begin{center}
\centerline{\epsfig{file=ruggfig2.ps,bbllx=55pt,bblly=30pt,bburx=535pt,bbury=850pt,
width=0.9\textwidth,height=0.85\textheight,clip=}}
\caption{Time evolution of the structure factor  for
 $\xi=2$, $d=6$, and  $\gamma=100$.
The projections on the planes  $k_{z}=0$ (left column) and $k_{y}=0$ 
(right column) are shown respectively for $t=0$ (top),
$t=2.5\times 10^{-2}$ (middle), when the maximum of
 $\sigma_{xy}$ is reached, 
and $t=2.0\times 10^{-1}$ (bottom).} 
\end{center}
\label{ruggfig_2}
\end{figure}
\newpage

\begin{figure}
\epsfig{file=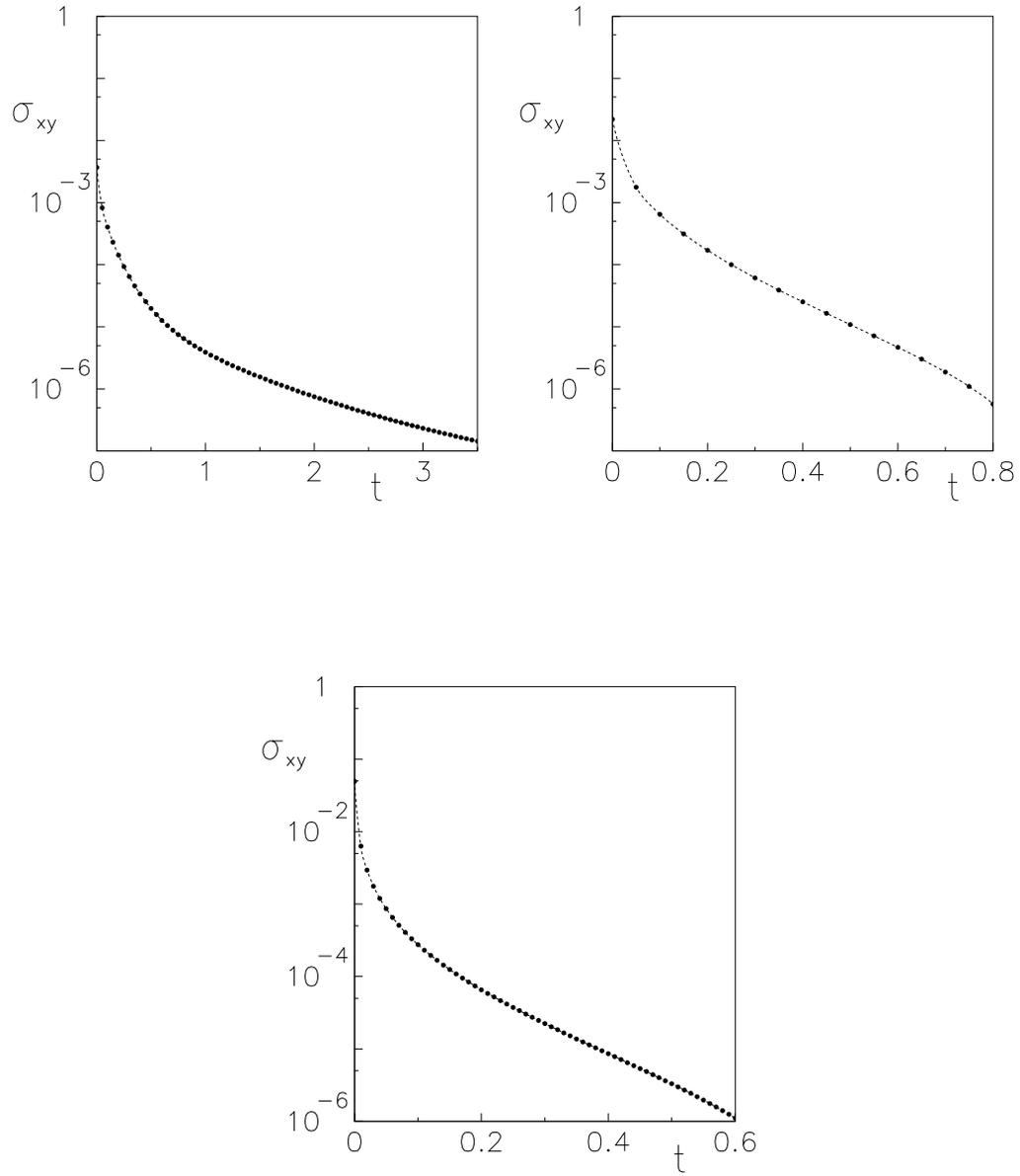,bbllx=90pt,bblly=221pt,bburx=509pt,bbury=700pt}
\label{fig_10}
\caption{Relaxation of the stress $\sigma_{xy}$ 
for $\xi=2$, $d=6$,
 $\gamma=2$ (top-left), $\gamma=20$ (top-right) and $\gamma=100$
(bottom). }
\end{figure}
\newpage

\begin{figure}
\epsfig{file=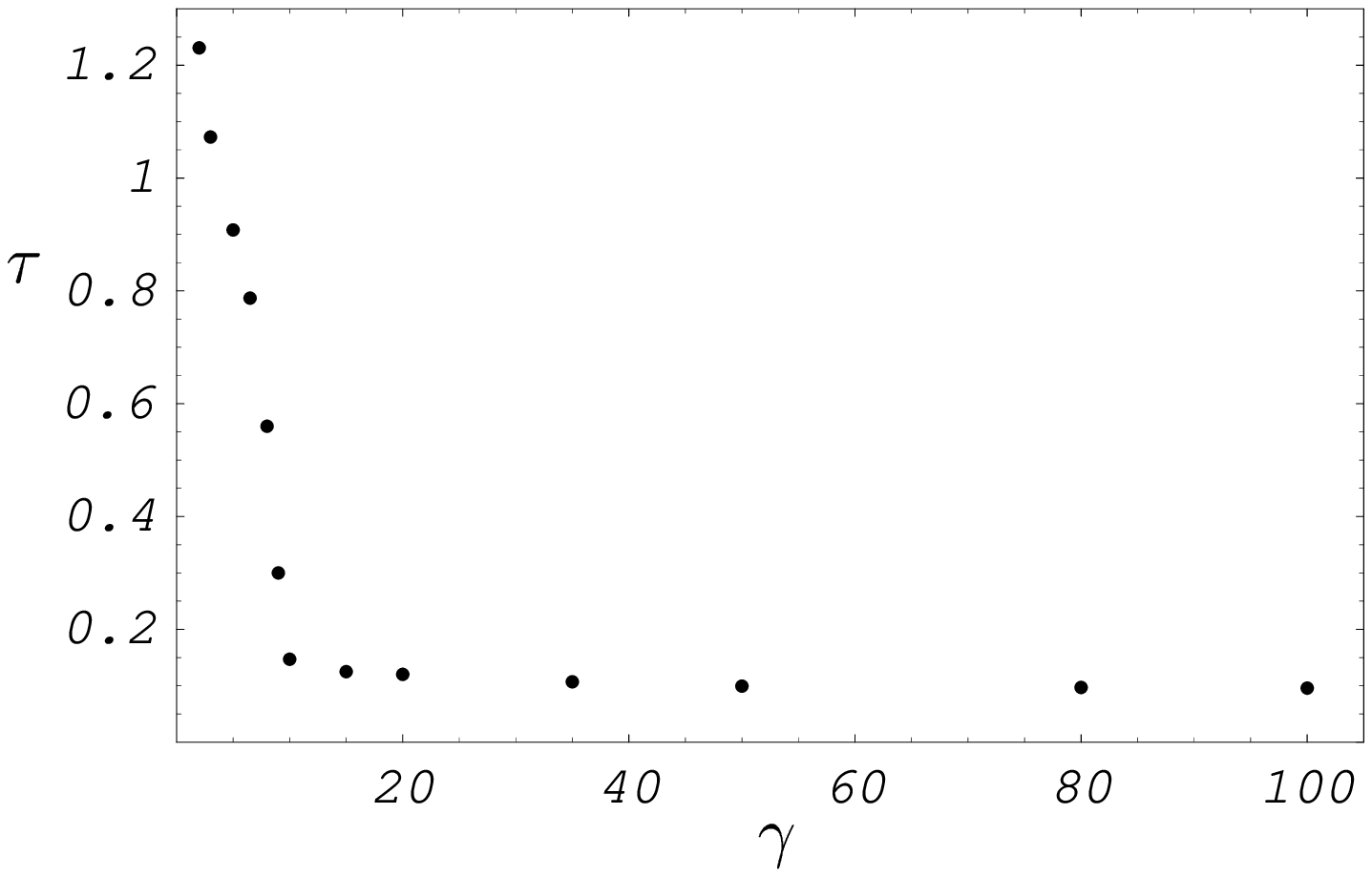,bbllx=72pt,bblly=300pt,bburx=538pt,bbury=603pt}
\label{fig_11}
\caption{Relaxation time $\tau$  as a function of the 
shear rate $\gamma$.}
\end{figure}


\begin{references}

\bibitem{larson}
See, e.g., R.G. Larson, {\it The structure and Rheology of Complex fluids}
(34~Oxford University Press, New York, 1999).

\bibitem{LJ73} 
Z. Laufer, H.L. Jalink, and A.J. Staverman,  Journal of Polymer Science
{\bf 11}, 3005 (1973).

\bibitem{GR}
For a review see, e.g., G. Gompper and M. Schick, 
{\it Self-assembling amphiphilic systems}
in {\it Phase 
Transitions and Critical Phenomena}, edited by C. Domb and J.L. Lebowitz
(Academic Press, New York, 1994), vol.16; 
K.A. Dawson in {\it Structure and Dynamics of Strongly Interacting
Colloids and Supramolecular Aggregates in Solution}, edited by 
S.-H.Chen at al. (Kluwer Academic Publ. 1992).

\bibitem{TS}
M. Teubner and R. Strey, J. Chem. Phys. {\bf 87}, 3195 (1987).

\bibitem{W}
B. Widom,  J. Chem. Phys. {\bf 90}, 2437 (1989).

\bibitem{col}
 A. Cappi, P. Colangelo, G. Gonnella, and A. Maritan,
Nucl. Phys. B {\bf 370}, 659 (1992);
 P. Colangelo, G. Gonnella, and A. Maritan,
Phys. Rev. E {\bf 47},  411  (1993); 
G. Gonnella and  J.M.J. van Leuween,
Phys. Rev. E, {\bf 52},  63 (1995).


\bibitem{Daw1}
G. P\"atzold and K. Dawson, Phys. Rev. E {\bf 54}, 1669 (1996);
G. P\"atzold and K. Dawson, J. Chem. Phys. {\bf 104}, 5932 (1996).

\bibitem{Sup}
F. Corberi, G. Gonnella, and D. Suppa,  Phys. Rev. E {\bf 63}, 040501(R).

\bibitem{OK}
A. Onuki and K. Kawasaki, Ann. Phys. (N.Y.) {\bf 121}, 456 (1979).

\bibitem{Fre}
G.H. Fredrickson, J. Chem. Phys. {\bf 85},  5306 (1986).

\bibitem{FL}
G.H. Fredrickson and R.G. Larson,
  J. Chem. Phys. {\bf 86}, 1553 (1987).

\bibitem{O87}
A. Onuki, J. Chem. Phys. {\bf 87}, 3692 (1987).

\bibitem{cates} 
M.E. Cates and S.T. Milner,  Phys. Rev. Lett. {\bf 62}, 1856 (1989).

\bibitem{HM}
C. Huang and M.  Muthukumar, J. Chem. Phys. {\bf 107}, 5561 (1997).

\bibitem {noi}
F. Corberi, G. Gonnella, and A. Lamura, Phys. Rev. Lett. 
{\bf 81}, 3852 (1998); Phys. Rev. Lett. {\bf 83}, 4057 (1999);

\bibitem{On97}
A. Onuki, J. Phys.: Condens. Matter {\bf 9}, 6119 (1997).

\bibitem{nota}
The approximation corresponds to take the limit 
$N \rightarrow  \infty $ in a model where the field $\phi$ 
is generalized to have $N$ components. This approximation 
differs only by a factor 1/3 multiplying $\lambda$ 
from the one-loop calculation of \cite{Daw1} and, for the purposes
of this paper, is equivalent to the one-loop approximation.


\end{references}
\end{document}